\documentstyle[epsf,twocolumn,aps]{revtex}
\begin{document}
\draft
\voffset 2.cm
\title{Multipositronic systems}

\author{K\'alm\'an Varga
\footnote{On leave of absence: 
Institute for Nuclear Research of the Hungarian Academy 
              of Sciences (ATOMKI), 4000 Debrecen, Hungary PO BOX 51}}
\address{Physics Division, Argonne National Laboratory, Argonne,
              Illinois 60439, USA}

\maketitle

\vspace{1cm}

\begin{abstract}
The stability and structure of systems comprising a negative ion
and positrons are investigated by the stochastic variational
method. It is shown that the H$^-$ and the Li$^-$ ions can bind
not only one but two positrons. The binding energies of these
double positronic atoms E(H$^-$,e$^+$,e$^+$)=0.57eV and 
E(Li$^-$,e$^+$,e$^+$)=0.15eV are somewhat smaller than those of their
single positronic counterparts (E(HPs)=1.06eV and 
E(LiPs)=0.32 eV). We have also found that a Ps$^-$, Ps$^-$ and 
a proton form a bound system. 
\end{abstract}
\narrowtext

The many-body problem is conceptually simple and well defined in 
atomic physics: indistinguishable fermions (electrons) interact 
via the Coulomb potential in the external Coulomb 
field of the atomic nuclei. The solution of this many-body problem 
is  very difficult, because in addition to the direct interaction 
between the electrons, their indistinguishability brings an exchange 
correlation into effect. Despite the complexity enormous progress 
has been made in this field which has been rapidly developing ever 
since the birth of quantum mechanics. The calculations have 
been mainly focused on systems (atoms and molecules) where fast electronic 
motion takes place in the field of slowly moving heavy positive charges.
\par\indent
Much less is known about systems which contain  positively 
and negatively charged particles of equal or nearly equal masses. 
The simplest examples of these systems are the Positronium ion $(e^+,e^-,e^-)$
(predicted by Wheeler \cite{wheeler} in 1946, experimentally observed by Mills
\cite{mills} in 1981), the Ps$_2$ molecule $(e^+,e^+,e^-,e^-)$ 
(predicted by Hylleraas and  Ore \cite{ore} in 1947, not observed yet 
in nature), or
the HPs molecule (indirectly observed, see \cite{Schr}). 
These systems have been extensively studied by various theoretical methods
in the last few years \cite{theoref}. The existence of these
small systems makes theorists curious as to whether (similarly to molecules)
larger stable systems containing positrons can also be formed. One can ask
whether a system of $m$ electrons $n$ positrons (for example an 
$(3e^-,3e^+)$ system) is bound or whether a positron, 
a positronium, a Ps$^-$ ion or a Ps$_2$
molecule can attach itself to an atom or molecule. 
\par\indent
The theoretical description of such systems (let alone the prediction 
of their stability against autodissociation) is obviously very 
difficult. The difficulty can largely be attributed to the fact 
that the electron-electron and the electron-positron correlations 
are quite different due to the attraction and to absence of the 
Pauli-principle constraint in the latter case. The 
tiny binding energies of these loosely bound extended systems 
require highly accurate calculations. 
\par\indent
Recent calculations have given the very  surprising result that 
a positron can cling to a neutral atom \cite{rm1}. The simplest such 
positronic atom is the Lie$^+$.  The complexity of the calculation 
of its small binding energy is best illustrated by the fact 
that many otherwise successful methods had failed to predict  the existence of 
the bound state of this system \cite{ward,yoshida}. These calculations 
show that the energy of the Lie$^+$ is lower than that of the Li atom but 
the energy was not below the Li$^+$+Ps dissociation threshold. The first 
rigorous proof showing that the positron can attach itself to a Li atom 
was given by Ryzhikh and Mitroy \cite{rm1} by using the stochastic 
variational method (SVM) \cite{kukulin,book}. This finding has been later 
confirmed by different theoretical approaches \cite{cdlin,dario,stras}. 
Other atoms (e.g. Be,Na,Mg,Cu,Zn and Ag) has also been found to be capable 
of binding a positron \cite{rmv1,rmv2,rm2}.
\par\indent
There is an  other family of positronic atoms which  are formed 
when positronium is attached to an atom. The possibility of the 
existence of such systems is more obvious: removing the positron 
leaves behind a negatively charged ion so one can understand how
the positron becomes bound. The simplest example of such system is the HPs 
molecule which has been the subject of numerous theoretical investigations 
and has been experimentally observed as well \cite{Schr}. Another examples are 
the LiPs, NaPs and KPs atoms. The LiPs has been described by several
microscopic methods \cite{dario,rmv1}, while the other two bound 
systems have been
predicted by a semi-microscopic model \cite{rmv2,rm2}.
\par\indent
In this paper we explore the possibility of the formation of stable 
atoms/ions containing two or more positrons. The simplest known 
example for such 
system is the Ps$_2$ molecule. The study is inspired by the speculation
that if a neutral atom can bind a positron then it may even be able
to bind a positively charged Ps$^+$=(e$^+$,e$^+$,e$^-$) ion. This motivation
can also be phrased in an other way: If  positronium can bind itself to 
a neutral atom ``A'' forming a neutral system ``APs'' then can we attach 
a positron into APs? 
\par\indent
The stochastic variational
method systematically improves the correlation functions 
between the particles and it is especially suitable to
solve Coulombic few-body problems. The method has been tested on a number
of problems in different fields of physics and it has been proved
to be highly accurate and reliable \cite{book,prl}. The present 
study is restricted to states with total orbital angular momentum L=0 
and the following trial function is assumed
\begin{equation}
\Psi={\cal A}\lbrace {\rm e}^{-{1\over 2}{\bf x}A{\bf x}}\chi_{SM_S}\rbrace ,
\end{equation}
where ${\bf x}=({\bf x}_1,...{\bf x}_{N-1})$ is a set of relative 
coordinates, $\chi_{SM_S}$ is the spin function, and $A$ is a matrix 
of nonlinear variational parameters. The nonlinear parameters are optimized by 
the stochastic variational method through a trial and error procedure.
The details can be found in Ref. \cite{book}.
\par\indent
This trial function includes  explicit ${\rm exp}(-\alpha r^2_{ij})$ 
correlation factors between the particles and it
gives very accurate solutions provided that the nonlinear parameters
(in the exponents) are properly optimized. As the number of parameters
for a typical system is at least a few thousands a direct search for the
optimal values is out of question. The stochastic variational method
sets up a basis successively enlarging the model space by including
the optimal trial functions. This basis was systematically improved 
by a refining procedure: The basis states were replaced  by randomly chosen 
states which lower the energy. The energy found in this variational 
procedure converges to the upper bound of the exact ground state energy 
of the system. The Correlated Gaussians offer computational 
advantage: fast analytical evaluation of the matrix elements and 
good approximation to various wave functions.  They also have 
well-kown drawbacks such as their slow convergence (compared to exponential 
functions) and the fact that they do not satisfy the cusp condition. 
\par\indent
The simplest (A,Ps$^+$)  is the (H,Ps$^+$)=(p,e$^-$,e$^-$,e$^+$,e$^+$)
system. This five-body system can dissociate into H+Ps$^+$, p+Ps$_2$
or HPs+e$^+$, the lowest dissociation thresholds are shown in Fig. 1.
To validate the method we have calculated the energies of the Ps$_2$
and HPs molecules (see Table I.). The SVM significantly improved
the theoretical values of the binding energies of these systems.
Our calculation shows that the energy of the (H,Ps$^+$)
is below the dissociation threshold and forms an electronically stable
system. The H$^-$ ion can bind not only one but two positrons. The binding
energy of (H,Ps$^+$)=HPse$^+$ (0.021 a.u) is comparable to that 
of HPs (0.039 a.u.). 
The convergence of the energy as a function of the basis dimension is 
shown in Table I.
\par\indent
The HPse$^+$ system can be also viewed as a bound system of a proton and
a Ps$_2$ molecule. The Ps$_2$ molecule cannot bind an extra electron or
positron because of the Pauli principle. Our calculations show that the
Ps$_2$ can bind a charged particle if it is distinguishable from the
electron and the positron. The binding energy of a five-body system
Ps$_2$+x$^+$=(e$^+$,e$^+$,e$^-$,e$^-$,x$^+$) consisting a hypotetical
``x'' particle is bound for any $0\le m_e/m_x \le 1$ mass ratio. This 
has been checked by calculating the binding energy of that system
for several different $m_x$ masses ($m_x=10^{50},100,10,8,6,4,2,1$ 
in units of m$_e$).
So while the (e$^+$,e$^+$,e$^+$,e$^-$,e$^-$) is unbound the Ps$_2$ can bind
any charged particle, e.g. $\mu^+$ or p$^+$, because the Pauli principle
does not restrict the motion of fifth particle in that case.
\par\indent
Some of the properties of these systems are shown in Table II.
It is intriguing to compare the relative distances between the particles
in HPs and HPse$^+$. The electron-nucleus or electron-electron
relative distances are almost the same in the two systems. The average
nucleus-positron distance, however is substantially
larger in HPse$^+$.  An other interesting property
is that the relative distance between positrons is about twice that 
between electrons. All these facts suggest that a possible geometrical 
picture of the HPse$^+$ looks as an isosceles triangle formed by the two
positrons and the proton and the two electrons are moving between the 
positive charges. The two positrons are placed on the vertices of the
baseline of the triangle, and this baseline is so long that the system
almost looks like as a linear chain. 
The HPse$^+$ is somewhat related to H$_3^+$. In H$_3^+$ three protons 
and two electrons form a very stable system, where the three protons are 
at the vertices of an equilateral triangle. By changing the mass of 
two of the positive 
charges this equilateral triangle is changed to an isosceles  triangle,
and in the positronic limit it looks like a linear chain.
\par\indent
The Li atom can bind a positron or a positronium forming an
electronically stable Lie$^+$ or LiPs \cite{rm1}. The binding energy 
of the Lie$^+$ is very small and it can be best viewed as a 
positronium orbiting around a Li$^+$ core. In our calculation we replace the 
positron with a Ps$_2$ ion and try to determine the binding energy.
In this case we have six active particles, four electrons and two 
positrons. This system has various different dissociation channels 
(see Fig. 2). The calculated energies of the relevant subsystems
are listed in Table III.
\par\indent
Our calculation shows that the Li can bind a Ps$^+$ ion to form an
electronically stable LiPse$^+$. The calculated binding energy might not be
very accurate due to the complexity of the system, but it is definitely 
below the lowest threshold (see Fig. 2). The convergence of the binding energy 
is shown in Table I. Further increase of the basis size would improve the 
ground state energy.
\par\indent
This system, again, can be viewed
in different ways. One can say that a Li atom can bind a Ps$^+$ ion, or a 
Li$^-$ ion is able to bind two positrons or the Ps$_2$ molecule can attach 
itself to a Li$^+$ ion. 
\par\indent
The relative distances between the particles in LiPse$^+$ are shown 
in Table III.
The average distance between the nucleus and the positron or between a 
positron and an electron is larger than that in LiPs but smaller than
what one can find in Lie$^+$. This would suggest a picture of LiPse$^+$ as a
Li$^+$ core with an orbiting Ps$_2$ molecule.
\par\indent
These systems are electronically stable but the positron electron 
pair can annihilate by emitting two photons. The annihilation rate is 
proportional to the probability of finding an electron and a positron 
at the same position in a spin singlet state (see eq. (21) in \cite{rmv2}).
The expectation values of the positron-electron delta functions
($\delta_{e^+e^-}=
\langle\Psi\vert\delta({\bf r}_{e^-}-{\bf r}_{e^+})\vert\Psi\rangle$)
are $1.4\times 10^{-2}$, $1.1\times 10^{-2}$ and $1.1\times 10^{-2}$,
for Lie$^+$, LiPs and LiPse$^+$.
 Due to the possible inaccuracy of the energy and
wave function of the LiPse$^+$ system the annihilation rate should be
considered as a qualitative estimate and it is about
$\Gamma_{2\gamma}=4.4 \times 10^9$ sec$^{-1}$.
\par\indent
The (H$^-$,e$^+$,e$^+$) is a positively charged system so one may
try to add one more electron to see if it remains stable. The 
convergence of the energy is shown in Fig. 3. The energy of the system
slowly converges to the lowest (HPs+Ps)  threshold and the size of the 
system continuously increases showing that this system is unlikely to be 
bound. Surprisingly, however, by adding two electrons to (H$^-$,e$^+$,e$^+$) 
one gets a bound system as shown in Fig. 3. This system ``H$^-$Ps$_2$''
contains a proton, two positrons and four electrons, and can also be 
considered as a three-body system of a proton a Ps$^-$ and a Ps$^-$ 
ion as an analogy
of the H$^-$ ion (where the electrons are replaced by the composite
Ps$^-$ ions). The convergence of the energy is slow and the calculation of
a more accurate binding energy would require a considerably larger basis
dimension (see Table I.)
\par\indent
We have shown, for the first time, that neutral atoms can bind not only a
single positron but a more complex positive charge, the Ps$^+$ ion as well.
Besides the two cases (HPse$^+$ and LiPse$^+$) it is quite possible that
other systems can be also bound. Although the investigation of larger 
systems is beyond the scope of the present method, other approaches (like
QMC \cite{dario,o2} or Fixed core SVM \cite{rmv2}) might  be used
to study the possible bound state of Ps$^+$ (or two positrons) 
with larger atoms/ions. Examples are (1) the recent QMC study of  
positronic water \cite{o2} and a new study with the Fixed core SVM which 
confirms the existence of the LiPse$^+$ and shows that a larger 
ion (Na$^+$) \cite{priv} can also bind a Ps$_2$ molecule.
\par\indent
The investigation of these exotic systems are very important from theoretical 
point of view. These systems serve as test grounds for new methods:
They provide a special environment where not only the 
electron-electron but other interleptonic correlations are also 
important. 
\par\indent
While the chance of experimental observation of these systems is even
more challenging than those of the positronic atoms \cite{rm3}, 
some of the properties of positronic systems can be affected by these 
bound states and the theoretical prediction of their existence might 
be very useful. 
\par\indent
Systems, similar to (p$^+$,e$^+$,e$^+$,e$^-$,e$^-$) might exist
in semiconductors. Both the charged exciton (system of two electrons and 
a hole, akin to Ps$^-$) and the biexciton (two electrons and two holes,
similar to Ps$_2$) have been experimentally observed \cite{chex,biex}. 
Larger systems of ``multiexcitons'' (system of several electron-hole pairs)
have also been observed \cite{mex1,mex2}. These systems are of 
course different from the electron-positron systems
because the electron-hole mass ratio ($\sigma=m_e/m_h$) differs from 
unity and also because there is no annihilation so their observation 
might be easier. The stability for electrons and positrons indicates 
the stability for systems with slightly different mass ratios.
The present study might give a hint for the existence 
of similar systems in semiconductors as well.
In GaAs, for example, there are heavy holes ($\sigma=0.196$) and light
holes ($\sigma=0.707$). A system similar to (p$^+$,e$^+$,e$^+$,e$^-$,e$^-$)
would comprise two electrons, a heavy, and two light holes. 
 
\par\indent 
This work  was supported by the U. S. Department of Energy, Nuclear 
Physics Division, under contract No. W-31-109-ENG-39 and OTKA grant 
No. T029003 (Hungary).

\begin{table}
\caption{The convergence of the total energy (E) and the 
energy relative to the lowest threshold ($\epsilon$) 
as a function of basis size. The energy value 
in parenthesis (below the name of the system) is the 
lowest dissociation threshold. Atomic units are used. Infinite mass
was used for the proton and Li nucleus.}
\begin{tabular}{cccc}
system   & Basis size&   E      &   $\epsilon$      \\
HPse$^+$     & 100       &$-0.809371$&     0.02017    \\
($-$0.78919) & 200       &$-0.809993$&     0.02080    \\
             & 400       &$-0.810152$&     0.02096    \\
\hline
LiPse$^+$    & 100       &$-7.79502$&      unbound    \\
($-$7.7959) & 200       &$-7.79811$&      0.00219    \\
            & 400       &$-7.80212$&      0.00620    \\
            & 800       &$-7.80510$&      0.00918    \\
\hline
(p$^+$,2e$^+$,4e$^-$)        
            & 100       &$-1.02981$&      unbound    \\
($-$1.0512)
            & 200       &$-1.04409$&      unbound    \\
            & 400       &$-1.05077$&      unbound    \\
            & 800       &$-1.05542$&      0.00423    \\
\end{tabular}
\end{table}
\begin{table}
\caption{Properties of Coulombic few-body systems. 
The  particle ``x'' is distinguishable from both 
the electron and the positron, but is has the same 
mass as the electron. Atomic units are used.}
\begin{tabular}{lcccccc}
system&Energy&
$\langle r_{e^-p}^2 \rangle$ &                      
$\langle r_{e^+p}^2 \rangle $&                     
$\langle r_{e^-e^-}^2 \rangle$&                       
$\langle r_{e^-e^+}^2 \rangle$&                       
$\langle r_{e^+e^+}^2 \rangle$ \\                       
HPs&$-0.78919$&7.81&16.25&15.87&15.58& \\
Ps$_2$&$-0.51600$&&&46.37&29.11&46.37 \\
HPse$^+$&$-0.81015$&7.49&31.84&15.14&33.71&65.40\\
x$^+$Ps$_2$&$-0.55647$&33.48&52.11&36.24&33.46&52.21\\
\end{tabular}
\end{table}
\begin{table}
\caption{Total energies and expectation values of various operators
in double positronic Li. Some other species are included 
for comparison. $r_{e^+}$ and $r_{e^-}$ are the distances between
the nucleus and the positron and the nucleus and the electron, 
respectively. Atomic units are used. }
\begin{tabular}{lcccccc}
system&Energy&
$\langle r_{e^+} \rangle $&                     
$\langle r_{e^-} \rangle $&                      
$\langle r_{e^-e^+} \rangle$&                       
$\langle r_{e^-e^-} \rangle$&                       
$\langle r_{e^+e^+} \rangle$ \\                      
Lie$^+$ &$-$7.5323&10.03& 3.44& 7.83& 6.43&        \\
LiPs    &$-$7.7397& 6.32& 2.82& 5.56& 4.75&        \\
LiPse$^+$&$-$7.8051& 7.51& 3.30& 6.12& 5.71&6.15    \\
\end{tabular}
\end{table}

\begin{figure}
\caption{Energy levels of the HPse$^+$ and the HPs+e$^+$ and H+PS$^+$
dissociation channels.}
\epsfxsize=3.cm
\epsfbox{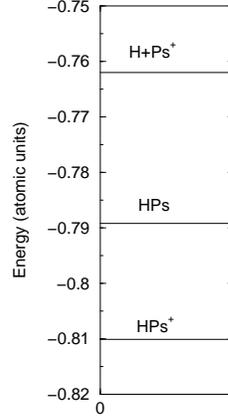}
\end{figure}

\begin{figure}
\caption{Energy levels of LiPse$^+$ and its most relevant dissociation 
channels.}
\epsfxsize=7.cm
\epsfbox{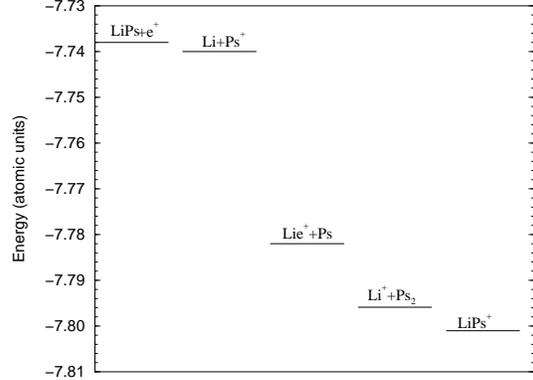}
\end{figure}

\begin{figure}
\caption{Convergence of the energy of the 
(p$^+$,e$^+$,e$^+$,e$^-$,e$^-$,e$^-$) and 
(p$^+$,e$^+$,e$^+$,e$^-$,e$^-$,e$^-$,e$^-$)
systems. The dotted line is the HPs+Ps, the dashed line is the H$^-$+Ps$_2$,
the long dashed line is the HPs+Ps$^-$ threshold.}
\epsfxsize=8.cm
\epsfbox{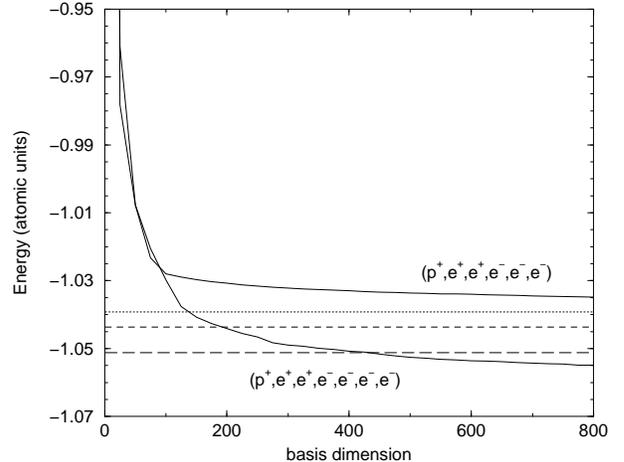}
\end{figure}

\begin{thebibliography}{99}
\bibitem{wheeler} J. A. Wheeler, Ann. N. Y. Acad. Sci. {\bf 48}, 219 (1946).
\bibitem{mills} A. P. Mills, Phys. Rev. Lett {\bf 46}, 717 (1981).
\bibitem{ore} E. A. Hylleraas and  A. Ore, Phys. Rev. {\bf 71}, 493 (1947).
\bibitem{Schr} D. M. Schrader, F. M. Jacobsen, N. P. Fradsen and U. Mikkelsen,
Phys. Rev. Lett. {\bf 69}, 57 (1992).
\bibitem{theoref} For example see the references in 
D. B. Kinghorn and R. D. Poshusta, Phys. Rev. A {\bf 47}, 3671 (1993), 
and in Y. K. Ho, Phys. Rev. A {\bf 48},4789 (1993).
\bibitem{rm1} G. Ryzhikh and J. Mitroy, Phys. Rev. Lett. {\bf 79} 
4124 (1998).
\bibitem{ward} S.J. Ward, M. Horbatsch, R. P. McEachran and A. D. Stauffer,
J. Phys. B: At.Mol. Opt. Phys. {\bf 22} 3763 (1998).
\bibitem{yoshida} T. Yoshida and G. Miyako, Phys. Rev. A {\bf 54} 4571 
(1996).
\bibitem{kukulin} V. I. Kukulin and V. M. Krasnopolsky, J. Phys. G3, 795 
(1977).
\bibitem{book} Y. Suzuki and K. Varga, Stochastic variational approach 
to quantum mechanical few-body problems, Springer-Verlag, 1998. 
\bibitem{cdlin} J. Yuan, B. D. Esry, T. Morishita and C. D. Lin,
Phys. Rev. A{\bf 58}, R4 (1998).
\bibitem{dario} D. Bressanini, M. Mella and G. Morosi, J. Chem. Phys. 
108, 4756 (1998).
\bibitem{stras} K. Strasburger and H. Chojnacki, J. Chem. Phys. 
{\bf 108} 3218 (1998).
\bibitem{rmv1} G. Ryzhikh, J. Mitroy and K. Varga,
J. Phys. B: At.Mol. Opt. Phys. {\bf 31} L265 (1998).
\bibitem{rmv2} G. Ryzhikh, J. Mitroy and K. Varga,
J. Phys. B: At.Mol. Opt. Phys. {\bf 31} 3965 (1998).
 \bibitem{rm2} G. Ryzhikh and J. Mitroy
J. Phys. B: At.Mol. Opt. Phys. {\bf 32} 1375 (1999).
\bibitem{prl} K. Varga, J. Usukura, and Y. Suzuki, Phys. Rev. Lett. {\bf 80}, 
1876 (1998).
\bibitem{o2} N. Jiang and D. M. Schrader, Phys.Rev.Lett {\bf 81} 5113 (1998).
\bibitem{rm3} J. Mitroy and G. Ryzhikh, J. Phys. B: At.Mol. Opt. Phys.
{\bf 32} L111 (1999).
\bibitem{priv} J. Mitroy and G. Ryzhikh, J. Phys. B: At.Mol. Opt. Phys. 
{\bf 32} L621 (1999).
\bibitem{chex} G. Finkelstein, H. Shtrikman and I. Bar-Joseph,
Phys. Rev. Lett. {\bf 74}, 976 (1995).
\bibitem{biex} D. Birkedal, J. Singh, V. G. Lyssenko, J. Erland, and
J. M. Hvam, Phys. Rev. Lett. {\bf 76}, 672 (1996).
\bibitem{mex1} A. G. Steele, W. G. McMullen, and M. L. W Thewalt,
Phys. Rev. Lett. {\bf 59} 2899 , (1987).
\bibitem{mex2} M. Bayer et. al. Phys. Rev. B{\bf 58} 4740 (1998).

\end{thebibliography}
\end{document}